# Museum: Multidimensional web page segment evaluation model

K.S. Kuppusamy and G.Aghila

**Abstract**— The evaluation of a web page with respect to a query is a vital task in the web information retrieval domain. This paper proposes the evaluation of a web page as a bottom-up process from the segment level to the page level. A model for evaluating the relevancy is proposed incorporating six different dimensions. An algorithm for evaluating the segments of a web page, using the above mentioned six dimensions is proposed. The benefits of fine-graining the evaluation process to the segment level instead of the page level are explored. The proposed model can be incorporated for various tasks like web page personalization, result re-ranking, mobile device page rendering etc.

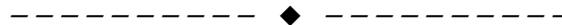

## 1 INTRODUCTION

The web search engines have become the pivotal axis of web experience for all kinds of users. More often than not web pages are accessed by clicking one of the result links from a search engine's output against a query entered by the user, rather than directly entering the address of the page (URL) in the browser.

In web search engines, the relevancy of a page with respect to a query is calculated by considering the entire page as an atomic unit. This paper proposes a model where the evaluation of a query against a page is done following the page segmentation process. Instead of calculating the weight of the whole page against a query, the page would be segmented into semantically and visually relevant components. The weight of the individual components can be calculated separately. While performing this calculation a set of weight coefficients, which are described later in this paper, are incorporated to make the calculation process efficient and relevant.

The calculation of the significance of a page against a query by splitting the page into multiple segments provides finer control over the evaluation process. This would emphasize the fact that different segments of page would contribute in a variable manner to the relevance of a page for a query instead of treating all components at the same weight.

The segment evaluation is carried out by following a six dimensional approach which facilitates the identification of importance of a segment against a query from multiple perspectives. The aggregation of weight associated with each of the segment is done to calculate the overall weight of the page against a query.

In the information retrieval domain the relevance of query against a page differs with respect to the user who has entered the query. The same query entered by different users need to fetch a different result set based on the profile of the user who has entered the query. Profile of the user is maintained to incorporate user specific weight in to the evaluation process.

This paper proposes the idea of combining the segmentation and temporal dimension in the evaluation process. The incorporation of segmentation of page in to smaller units and repeating it for the snapshots gathered at different time intervals is also considered as an evaluation factor in this research work.

The objectives of this research are the following:
1) Proposal of a multi-dimensional web page segment evaluation model.
2) Exploring the benefits achieved by fine-graining the evaluation process to the segment level instead of page level.

The rest of this paper is organized as follows: The state-of-art works that have been carried out in this field are discussed in section 2. The proposed model is explained in section 3. The algorithm is given in section 4. In section 5, the conclusions and future directions for this research work are listed.

## 2 RELATED WORKS

This section would highlight the state-of-art works that have been carried-out related to the theme of this paper.

The web page segmentation has been explored by various researchers. There exist various approaches to segment a web page. Cao et al [1] has proposed a segmentation method based on image processing techniques. The web page is considered as an image and features are used to segment the web page. Kohlschütter et al [2] has ap-

---

- *K.S. Kuppusamy is with the Department of Computer Science, School of Engineering and Technology, Pondicherry University, Pondicherry 605014.*
- *G. Aghila is with Department of Computer Science, School of Engineering and Technology, Pondicherry University, Pondicherry 605014.*

plied the text density of sections of the page to carry out the segmentation process. A graph theory based segmentation approach is explained by Deepayan Chakrabarti et al [3].

The Vision based Page segmentation (VIPS) process explained by D.Cai et al [4] provides the segmentation approach based on visual features. Since this approach is closer to the human perusal of a web page, the segmentation for this work has been carried out with this approach.

Learning the imporatance of the page segments is explored by Ruihua Song et al [5]. It has been indicated that differentiating noisy segments from the informative ones can be a very handy mechanism in better managing the web page for mining and other processes.

Lin et al [6] has explored a table based approach to identify informative portions from a web page. The segment handling in dynamic web pages is illustrated by Ramaswamy et al [7].

A unified approach to document similarity search using manifold-ranking of blocks has been proposed by Xiaojun Wan et al [8].

## 2 THE MUSEUM MODEL

The proposed model is termed as "Museum". This term is an acronym for **Mu**ltidimensional **Se**gment evaluation **m**odel. At the same time the literal meaning of the term Museum is also relevant to the theme of the paper. Museum is the place where a wide collection of valuable resources are displayed for user viewing. Since the theme of this paper is information retrieval, the term Museum adds a special meaning in the current context.

In our approach we divide each page in to segments of semantically and visually related components. Each segment would be weighed against the query from different dimensions. So the web page can be divided into differ-segments

$$P_i = \{ s_1, s_2, s_3 ..., s_k \} \tag{1}$$

The segmentation of a page is done satisfying the following two rules:

**Rule 1**: During segmentation the components are selected such that they are non-overlapping.

$$\forall (s_i, s_j): s_i \cap s_j = NULL ; i,j = (1, k) \tag{2}$$

**Rule 2**: Segmentation incorporates all parts of the web page.

$$s_1 \cup s_2 \cup s_3 \cup .... s_k = P_i \tag{3}$$

$$P_i = \bigcup_{j=1}^{k} S_k \tag{4}$$

The weight associated with a segment $\omega(s_i)$ is determined by a multimodal estimation process. The $\omega(s_i)$ is

represented as:

$$\omega(s_i) = (F, E, L, V, R, M) \tag{5}$$

Where
F = Freshness weight Co-efficient
E = Theme Weight Co-efficient
L = Link Weight Co-efficient
V = Visual Weight Co-efficient
R = Profile Weight Co-efficient
M = Image Weight Co-efficient

**Freshness weight coefficient:** Freshness weight coefficient $\omega_F(s_i)$ is assigned a value of number of query terms matching the terms in the segment if and only if the intersection of previous snapshots in the page evolution track and the query terms is NULL.

$$\forall t_i \in T, q_i \in Q : |t_{i-1} \cap q_i| = NULL \tag{6}$$

The $\omega_F(s_i)$ is calculated as the sum of actual freshness weight coefficient and synonym freshness weight coefficient.

$$\omega_F(s_i) = \omega_{FA}(s_i) + \omega_{Fs}(s_i) \tag{7}$$

Where the actual freshness coefficient $\omega_{FA}(s_i)$ is calculated as:

$$\omega_{Fs}(s_i) = \begin{cases} \forall t_i \in T, q_i \in Q : |t_i \cap q_i \cap s_i| & if |t_{i-1} \cap q_i| > 0 \\ 0 & otherwise \end{cases} \tag{8}$$

The actual freshness weight coefficient $\omega_{FA}(s_i)$ is calculated as the number of query terms matching in the segment if first part of the condition is met.

The synonym freshness weight coefficient $\omega_{Fs}(s_i)$ is calculated as the number of synonyms of query terms matching in the segment, divided by two.

$$\omega_{Fs}(s_i) = \begin{cases} \forall t_i \in T, q \in Q : |t_i \cap syn(q) \cap syn(s_i)| / 2 & if |s_{i-1} \cap syn(q)| > 0 \\ 0 & otherwise \end{cases} \tag{9}$$

**Theme Weight coefficient :** The theme of a web page is represented by the title given to that web page. If the segment consists of terms from page title then the theme weight coefficient is assigned as the number of terms matching between the title and segment terms.

$$\omega_T(s_i) = \begin{cases} \forall s_i \in E, q_i \in Q : |s_i \cap q_i| + |syn(e_i) \cap syn(t_i)| / 2 & if |syn(s_i) \cap syn(e_i)| > 0 \\ 0 & otherwise \end{cases} \tag{10}$$

**Image Weight coefficient**: The image weight coefficient $\omega_M(s_i)$ is defined as the number of query terms that matches in the alt attribute of the image.

$$\omega_M(s_i) = \begin{cases} \forall m_i \in M, q_i \in Q : |s_i \cap (m_i)| + |s_i \cap ((m_i))| / 2 & if |s_i \cap ((m_i))| > 0 \\ 0 & otherwise \end{cases} \tag{11}$$

**Link Weight coefficient**: The link weight coefficient $\omega_L(s_i)$ is defined as the number of query terms appearing as links in the given segment $s_i$.

$$\omega_L(s_i) = \begin{cases} \forall l_i \in L, q_i \in Q : |s_i \cap l_i| + |syn(s_i) \cap syn(l_i)| / 2 & if |syn(s_i) \cap syn(l_i)| > 0 \\ 0 & otherwise \end{cases} \tag{12}$$

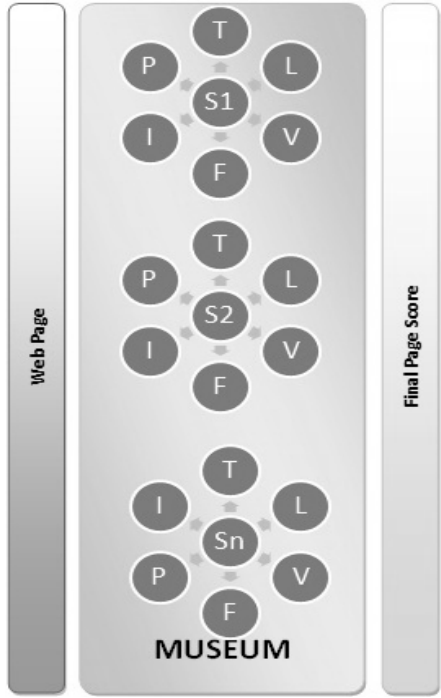

Fig. 1. The Museum Model

Where L is the link array in the segment $s_i$.

**Profile Weight coefficient**: The profile weight coefficient $\omega_P(s_i)$ is defined as the number of query terms that matches in the profile of the user.

$$\omega_R\big|s_i\big| = \begin{cases} \forall r_i \in R, q_i \in Q : |s_i \cap (r_i)| + |s_i \cap syn(r_i)| / 2 & if |syn(s_i) \cap syn(r_i)| > 0, \\ 0 & otherwise \end{cases} \quad (13)$$

**Visual Weight coefficient**: The visual weight coefficient $\omega_V(f_i)$ is defined as the number of query terms appearing in segment with specific visual markup. Each item in the visual weight array is given different weights

$$\omega_V = \big\{ [v_1, \omega_{V1}], [v_1, \omega_{V2}], \dots [v_1, \omega_{Vn}] \big\} \quad (14)$$

Where
$v_i$ is the visual markup item.
$\omega_{Vi}$ is the weight for the visual markup item $v_i$

$$\omega_V(f_i) = \begin{cases} \forall v_i \in V, q_i \in Q : |q_i \setminus v_i| * \omega_{vi} & if |v_i \cap q_i| > 0 \\ 0 & otherwise \end{cases} \quad (15)$$

The sum of all above mentioned coefficients would provide the total weight of the corresponding segment $s_i$.

$$\omega(s_i) = \omega_F(s_i) + \omega_B(s_i) + \omega_L(s_i) + \omega_V(s_i) + \omega_d(s_i) + \omega_R(s_i) \quad (16)$$

So the overall score of the page will be given by

$$\omega_p(P_i)^J = \frac{1}{y} \sum_{k=1}^{y} \omega(s_k) \quad (17)$$

Where y indicates the total number of segments in the web page.

The Museum model is illustrated in Fig 1. The three columns in the figure illustrate the input web page, the model components and the resultant score, from left to right. Each of the circular units specified in the middle column of the figure depicts the segments and it's six dimensions based evaluation.

## 3 THE SEGMENT WEIGHT ALGORITHM

The algorithm for evaluating the segments is as given below.

Algorithm SegmentWeight

Input: Page P,  Page Evolution Track PET , Query Q

Output : Segment Weight matrix (P, S)

Begin

identify the segment set S = {s1,s2 ….. sk }

initialize the visual feature array VF

for each segment sj in S

// Check if the segment is newly introduced and relevant to query

 if sj is a new segment, PET[sj] ∩ sj = NULL then

if sj ∩ {q1 , q2 , … qn } <> NULL then

set  fresh content flag  freshweight[sj] = | sj  ∩ {q1 , q2 , … qn }|

if sj  ∩ syn[{q1 , q2 , … qn }] then

set related fresh content flag freshcontent[fj] = | sj  ∩ syn[{q1 , q2 , … qn }] | / 2

// Boost the weight if segment contains elements in title

if (sj ∩ title[Pi]) then set theme flag,   themeweight[sj] = |sj ∩ title[Pi]|

// Boost the weight if links in segment contains query terms if ( link[sj] ∩ ({q1 , q2 , … qn } ∪ syn[{q1 , q2 , … qn }])) then set linkweight[sj] = | link[sj] ∩ ({q1 , q2 , … qn } ∪ syn[{q1 , q2 , … qn }])|

// Boost the weight based on visual features

for each vfr in VF

if (sj ∩ vfr) <> NULL

set visualweight[sj] ) = count(style({q1 , q2 , … qn }) in vfr)

// Boost the weight based on user profile

If (sj ∩ fetch profile keywords( ) ) then set profileweight(sj) = | sj ∩ fetch profile keywords( )|

// Boost the weight based on image attributes

If (sj ∩ fetch image alt ( ) ) then set imageweight(sj) = | sj ∩ fetch image alt ( )|

compute the consolidated segment weight (Pi, Sj) = freshweight[sj] + themeweight[sj] + linkweight[sj] + visualweight[sj] + profileweight(sj) + imageweight(sj)

End

## 5 CONCLUSIONS AND FUTURE DIRECTIONS

The proposed Museum model evaluates a web page based on six different dimensions. Finally the scores are merged to calculate the final score of the page. The benefits of proceeding with this apporch are as listed below:

- The local relevance of the segments can be utilized. This facilitates distinguishing normal segments from important ones.
- The incorporation user profile in the segment evaluation model provides personalized weight calculation.
- The inclusion of visual features inside a segment, finetunes the evaluation process.

The future directions for this work would include the following.

- Exploring the benefits of applying the Museum Model for search result reranking.
- Implementing the Museum Model to provide personalized page rendering.
- Analysing the suitability of Museum model for mobile devices page rendering.

## REFERENCES

[1] Cao, Jiuxin , Mao, Bo and Luo, Junzhou, 'A segmentation method for web page analysis using shrinking and dividing', International Journal of Parallel, Emergent and Distributed Systems, 25: 2, 93 — 104, 2010.

[2] Kohlschütter, C. and Nejdl, W. A densitometric approach to web page segmentation. In Proceeding of the 17th ACM Conference on information and Knowledge Management (Napa Valley, California, USA, October 26 - 30, 2008). CIKM '08. ACM, New York, NY, 1173-1182, 2008.

[3] Deepayan Chakrabarti , Ravi Kumar , Kunal Punera, A graph-theoretic approach to webpage segmentation, Proceeding of the 17th international conference on World Wide Web, April 21-25, Beijing, China, 2008.

[4] D. Cai, S. Yu, J. Wen, and W.-Y. Ma, VIPS: A vision-based page segmentation algorithm, Tech. Rep. MSR-TR-2003-79, 2003.

[5] Ruihua Song , Haifeng Liu , Ji-Rong Wen , Wei-Ying Ma, Learning block importance models for web pages, Proceedings of the 13th international conference on World Wide Web, May 17-20, New York, NY, USA, 2004.

[6] Lin, S.-H. and Ho, J.-M., *Discovering Informative Content Blocks from Web Documents*, in the proceedings of the ACM SIGKDD International Conference on Knowledge Discovery & Data Mining (SIGKDD'02), 2002.

[7] Ramaswamy, L., Iyengar, A., Liu, L., and Douglis, F. 2004. Automatic detection of fragments in dynamically generated web pages. In Proceedings of the 13th international Conference on World Wide Web , New York, NY, USA, May 17 - 20, 2004. WWW '04. ACM, New York, NY, 443-454, 2004.

[8] Xiaojun Wan, Jianwu Yang, Jianguo Xiao, Towards a unified approach to document similarity search using manifold-ranking of blocks, Information Processing & Management, Volume 44, Issue 3, May 2008, Pages 1032-1048, ISSN 0306-4573.

**K.S.Kuppusamy** is an Assistant Professor at Department of Computer Science, School of Engineering and Technology, Pondicherry University, Pondicherry, India.He has obtained his Masters degree in Computer Science and Information Technology from Madurai Kamaraj University. He is currently persuing his Ph.D in the field of Intelligent Information Management. His research interest includes Web Search Engines, Semantic Web.

**G. Aghila** is a Professor at Department of Computer Science, School of Engineering and Technology, Pondicherry University, Pondicherry, India. She has got a total of 20 years of teaching experience. She has graduatedfrom Anna University chennai, India. She has published nearly 40 research papers in web crawlers, ontology based information re-trieval. She is currently a supervisor guiding 8 Ph.D. scholars. She was in receipt of schrneiger award. She is an expert in ontology development. Her area of interest inlcude Intelligent Information Management, artificial intelligence, text mining and semantic web technologies